# On the concept of mass in relativity

Peter M. Brown, 786-C Washington Street, Haverhill, MA 01832
E-mail: pmb61@hotmail.com

Within the past fifteen years the use of the concept of *relativistic mass* has been on the decline and has been replaced by the concept of *proper mass (aka rest mass)* — simply referred to as *mass* and labeled *m* by its proponents. This decline in usage appears to be due to arguments presented in several journal articles over the last thirty-five years, as well as to standard practices in the field of particle physics. The aforementioned debate consists of arguments as to how the term "mass" should be defined to maximize logic as well as to be less confusing to the layman and people just starting to learn relativity. Lacking in the debate is a clear definition of all types of mass and all its usages in a wide variety of cases. The purpose in this article is to bring a unifying perspective to the subject. In doing so I will explore those things omitted from previous articles on this subject including the importance of point particles vs. extended objects; open vs. closed systems and gravitational mass. Although I argue for the usage of relativistic mass I do *not* argue that proper mass is not an important tool in relativistic dynamics.

I. INTRODUCTION

As early as 1968 (perhaps even earlier) physicists have been attempting to downplay the usage of the concept of the inertial mass (aka relativistic mass). Inertial is defined as the ratio of the magnitude of the momentum of a particle to the speed of the same particle, i.e. $m \equiv p/v$.[1] However more and more physicists are defining the term *mass* to mean what is often referred to as proper mass and labeled, *m*, and are leaving out or diminishing mention of relativistic mass. This decrease of usage can only serve to damage a full understanding of mass and relativity itself as will be shown below.

Since 1989 there have been several articles published on this topic [2-7], the main intent of these articles is to attempt to eradicate the concept of relativistic mass completely from newer publications of textbooks. For example, some of the articles state "[relativistic mass] deserves a much more subordinate status than it enjoys..." [1], "[relativistic mass] is of no legitimate use..." [2], "It is our duty -- as professional physicists -- to stop this process." [3], etc. in textbooks. It is also being said that it should only be used as an example of a historical archaic origin.", and "It is time for physics education community to abandon the notion of relativistic mass once and for all." [4]. This differing of opinions and attempts at prohibiting relativistic mass has taken the form of a lively and passionate debate in the physics literature. [2-7] The various masses that exist in classical mechanics and arguments given in the ongoing debate about relativistic mass are listed respectively in Tables 1 and 2 below. Those arguments in the "Pro-Relativistic Mass" column without a reference are presented in this article.

| Table 1 | |
|---|---|
| Concept of Mass | Definition of Concept of Mass |
| Proper Mass/Rest mass | If such a quantity exists in a given situation then it is that quantity for which there exists a unique number representing the Newtonian mass, regardless of the coordinate system used (invariant). |
| Inertial Mass | The ratio of momentum to speed, i.e. $m = p/v$. |
| Passive Gravitational Mass | The mass that gravity acts on. |
| Active Gravitational Mass | Mass that is the source of gravity. |

| Table 2 | |
|---|---|
| Con-Relativistic Mass | Pro-Relativistic Mass |
| 4-momentum is defined as "proper mass"×"4-velocity". Assigning mass a relativistic character obscures the physics. [1] | Arguing against relativistic mass might suggest to some who have not thought these things out that there are unresolved difficulties or commit a blunder whereas this is all about terminology. [4] |
| There is confusion between $m$ and $m_0$. [1] | $m = \gamma m_0$ is a useful heuristic concept. [4] |
| $m = E/c^2$ has no rational justification. [3] | Relativistic mass is the most consistent and understandable way to introduce many concepts of relativistic mass to students. [6] |
| Particle physicists only use the term *mass*. [3] | Force is *defined* as $\mathbf{F} = d\mathbf{p}/dt$ *not* as $\mathbf{F} = m\mathbf{a}$ [6] |
| $m$ confuses both teachers and students. [3] | The position vector for the center of mass is better understood when defined using relativistic mass rather than defined using only proper mass. As such the conservation of the center of mass theorem is easier to comprehend. |
| Can't cling onto the "Newtonian relation" $F = ma$. [2] | It is agreed by all that "light carries mass." It is not agreed that "light has mass" in all cases. It is easier to understand "light carries mass" when one can also say that "light *has* mass." |
| $E/c^2$ is not gravitational mass. The source of gravity is the energy-momentum tensor. The gravitational force in general relativity is *not* given by $\mathbf{F}_g = -\dfrac{GMm}{r^3}\mathbf{r}$. [3] | Light has passive gravitational mass and active gravitational mass. Each is identical to relativistic mass. Omitting this fact obscures general relativity since mass is the source of gravity. [11] |
| Einstein said that it is not good to introduce the quantity $m = \gamma m_0$. [3] | Any so-called "confusion" regarding relativistic mass can be eliminated in two ways; the relativity teacher explains both points of views and erroneous concepts are corrected. [6] |
| Texts are contradictory on what $m$ means and this confuses students. [3] | |
| Leading journals don't use $m = \gamma m_0$. [3] | |
| $m = \gamma m_0$ is archaic/old fashioned, a fiction which is best avoided. [9] | |
| Special relativity is a theory about geometry and from the modern, geometric viewpoint, the mass is invariant, not velocity dependant. [5] | |
| $m$ is total energy, not mass. [10] | |

Proponents of relativistic mass (the "pro-side") assert that the relativistic mass of an object can in all cases be specified when its speed and proper mass are given in the observer's frame of reference. Thus, it is held, relativistic mass is well defined by a single number in all cases. It will be shown below that this assertion is invalid in general. But it holds only in certain special circumstances such as isolated objects, complete systems, and point particles. An isolated object is an object that is self-contained in that it does not interact with anything external to the object. E.g. an ideal gas in a box is not an isolated system but the box containing it plus the gas is and isolated system. A complete system consists of all that constitutes the object and related fields and as such the field is included as being part of the object. When an object is not isolated then the mass of the object will be a function of the direction of momentum. Different directions of momentum will yield different masses.

According to the pro-side an object's inertia is manifested in its ability to resist changes in momentum. According to the "con-side" of this debate an object's inertia is manifested in its ability to resist changes in velocity and is complicated by the fact that generally cannot be represented by a single number. In relativistic dynamics the 3-force can be written as $\mathbf{f} = m_l \mathbf{a}_l + m_\perp \mathbf{a}_\perp$ where $\mathbf{a}_l$ is the longitudinal component of acceleration and $\mathbf{a}_\perp$ is the transverse component of acceleration. The quantities $m_l$ and $m_\perp$ are then referred to as the object's *longitudinal mass* and the *transverse mass*, respectively. Thus for relativistic speeds this ratio depends on the direction of acceleration and in general is not a single number but well defined by two numbers when the accelerated object is a single particle. The former definition utilizes Weyl's definition of mass ($m \equiv p/v$) while the later utilizes Mach's definition ($m \equiv F/a$).[8] It is Weyl's definition that is used in relativity on both sides of this debate and each is referred to as *inertial mass*. It appears that the only place $m \equiv F/a$ finds use is in arguments against relativistic mass.

As expected, proponents on each side of the debate assert that their definition is the "correct definition" (i.e. most logical and least misleading). Others and I disagree with the assertion that inertial mass should be defined by the relation $m \equiv F/a$.[6] This definition incorrectly assumes that $\mathbf{F} = m\mathbf{a}$ is always true. However it's not even generally true in Newtonian mechanics.

It is also asserted by the con-side that "mass," however it is defined, should be a geometric quantity due to the use of geometry to describe relativity. The geometric property that is spoken of is the magnitude of a particle's 4-momentum. However if this is the case then one should use the term *proper* to qualify this term. The magnitudes of several 4-vectors are often named in this way. Therefore the "geometric term" should be *proper mass* and not simply *mass*. It is also held by the con-side that "mass," however it is defined, cannot be thought of as gravitational mass and in doing so one is both mistaken as well as easily misled.[3] In this article I will outline these assertions below and bring to light any erroneous assumptions/arguments that have been made in the literature concerning this debate. A brief outline of the debate is as follows.

Some notation and terminology are defined in the appendix.

II. DEFINITION OF MASS

From here on the term *mass*, when used unqualified, will to refer to *relativistic mass* (aka *inertial mass*) and labeled $m$. Proper mass will be labeled $\square$. In most cases the proper mass of a particle equals the rest mass of the particle with the exceptions described below. The qualifier *proper* is used to denote something that is an *intrinsic property*. E.g. the proper mass of an electron is an intrinsic property of the electron. Within both Newtonian and relativistic mechanics there are three subdivisions in the concept of mass, which include, as Table 2 states, are

*inertial mass* – Mass which resists changes in momentum.

*passive gravitational mass* – Mass acted upon by gravity.

*active gravitational mass* – Mass that is the source of gravity.

The distinction between the later two has been left out of the mass debate entirely. This situation will be rectified in the sections below. It will be demonstrated in this article that each of these three quantities have the same numerical values and as thus all deserving of the name *mass*.

In this section I address only *inertial mass*. The inertial mass of a particle is the quantity $m$ such that in collisions with other particles in which $n$ free particles enter a collision and $p$ free particles exit the collision satisfies, the relation

(1) $$\sum_{i=1}^{n} m_i \mathbf{v}_i = \sum_{i=p+1}^{n+p} m_i \mathbf{v}_i$$

in all inertial frames of reference in flat spacetime. The quantity $\mathbf{p} \equiv m\mathbf{v}$ is then defined as the particle's *3-momentum*. We can loosely say that *mass is defined such that 3-momentum is conserved*. A tardyon particle (a particle for which $v < c$ is always true. E.g. an electron) moving with speed $v$ with respect to S will be a function of the particle's speed and can be shown to be related to ☐ by [12]

(2) $$m = m(v) = \gamma\mu, \quad \gamma \equiv \frac{dt}{d\tau} = \frac{1}{\sqrt{1-\beta^2}}, \quad \beta \equiv v/c$$

The equality $\gamma = 1/\sqrt{1-\beta^2}$ only holds in inertial frames of reference. Since ☐ ≡ $m(0)$ we see where the term "rest mass" derives its name. Eq. (2) may be expressed in terms of the *total inertial energy*, $E = mc^2$, (kinetic energy + rest energy, i.e. $E = K + ☐c^2$). The total energy $W$ has the value $W = K + ☐c^2 + V$ where $V$ is the particles potential energy. $W/c$ is the time component of the canonical momentum 1-form $\tilde{\pi}$. For a charged particle in an electromagnetic field the spatial components of $\tilde{\pi}$ are found to be $\pi_k = p_k + qA_k$. Notice that in general $\tilde{\pi} \neq \mathbf{P}$. If the particle is a luxon (a particle for which $v = c$ is always true, e.g. a proton) then the mass is given by $m = p/c$. This may seem strange to many readers. However, many relativity texts that utilize relativistic mass assign such a mass to photons. Even Einstein assigned a mass to radiation. [13-20] This is unavoidable since relativistic mass is defined through momentum and therefore anything that has momentum must, therefore, have mass. Note that since luxons have $E = pc$ this implies from the expression $\sqrt{E^2 - (pc)^2} = \mu c$ that $\mu = 0$. It is for this reason that it is said that the proper mass of a photon is zero even though the derivation of the equation from which this deduction is made was derived under the assumption that the particle was a tardyon. Since a photon can never be at rest this is the first good reason not to use the term rest mass when proper mass is more accurate.

Another definition of proper mass is given in terms of the conservation equation

$$(3) \quad \sum_{i=1}^{n} \mu_i \mathbf{U}_i = \sum_{i=p+1}^{n+p} \mu_i \mathbf{U}_i$$

The quantity $\mathbf{P} \equiv \mu \mathbf{U}$ was defined by Minkowski, who referred to $\mu$ (which he labeled $m$) as the *mechanical mass* and to $\mathbf{P}$ as the "momentum vector." [21] This definition fails if one or more of the particles are luxons and thus fails as a general definition. So as not to confuse $\mathbf{P}$ with $\mathbf{p}$ it has become customary to refer to $\mathbf{P}$ as the particle's *4-momentum*.

The components of the 4-momentum are given by

$$(4) \quad \mathbf{P} \equiv \mu \mathbf{U} = \mu \frac{d\mathbf{X}}{d\tau} = \mu \frac{d}{d\tau}(ct, x, y, z) = \mu\left(c\frac{dt}{d\tau}, \frac{dx}{d\tau}, \frac{dy}{d\tau}, \frac{dz}{d\tau}\right)$$
$$= \mu(c\gamma, \gamma v_x, \gamma v_y, \gamma v_z) = (\gamma\mu c, \gamma\mu v_x, \gamma\mu v_y, \gamma\mu v_z) = (\gamma\mu c, \mathbf{p})$$

Up to this point we have no reason to assume that the timelike component of $\mathbf{P}$ is related to energy as many physicists do. Since $P^0$ is directly proportional to mass for a particle in an inertial frame then this is the motivating reason for setting $E = cP^0$. However it will be shown below that the relation $E = mc^2$ holds only in coordinates corresponding to an inertial frame of reference. Components of 4-vectors should never be defined with a particular coordinate system in mind. Therefore the substitution $\gamma\mu c = mc$ should be made to obtain

$$(5) \quad \mathbf{P} = (mc, \mathbf{p})$$

Eq. (5) may also seem odd to some readers since the time component of $\mathbf{P}$ is often expressed as $P^0 = E/c$. However there are several texts where the authors prefer to define 4-momentum with $mc$ as the time component rather than $E/c$. [22-24] It is readily seen that the mass of a particle, as measured by an observer whose 4-

velocity is $\mathbf{U}_{obs}$ is proportional to the scalar product $\mathbf{U}_{obs}$ and $\mathbf{P}$, i.e. $m = \mathbf{P} \cdot \mathbf{U}_{obs}/c$. Thus mass is a covariant quantity since it depends on an observer and hence on a coordinate system. The inertial energy of a particle is best defined as proportional to the time component of $\tilde{\mathbf{p}}$, i.e. the 1-form corresponding to 4-momentum. This is consistent with the definition of total energy defined above for a charged particle in an electromagnetic field. This quantity is a constant when the 4-force on the particle is zero (e.g. when in free-fall in a gravitational field) and the gravitational potentials are time independent (i.e. the gravitational field is static). [25] That these definitions are the preferred ones follows from the fact that only in inertial frames are the time components of $\mathbf{P}$ and $\tilde{\mathbf{p}}$ proportional to each other and as such one should never define quantities which are coordinate dependant.

Let frame S and S′ be in standard configuration. The components of $\mathbf{P}$ as measured in S and S′ are related to each other by a Lorentz transformation corresponding to a boost in the $+x$ direction. If the spatial component of a 4-vector is conserved in all inertial frames of reference then it can be shown that the temporal component is also conserved in all inertial frames. [26] Since this the case for $\mathbf{P}$ then it follows that mass is conserved in all inertial frames under the conditions specified for Eq. (1). It should be noted that at no time was energy conservation involved and yet the conservation law for mass can be derived. This is one of the usefulness qualities of relativistic mass.

One of the main objections of those who reject this definition of mass claim that since relativistic mass is proportional to energy then one is merely giving energy a different name and there is no valid reason for doing so. There is a serious flaw in this argument. It assumes that all cases $E = mc^2$ but it is known that $E = mc^2$ holds only in very special cases such as that of isolated objects or point objects such as an electron. [27] Before we proceed the reader should recall the statement by Einstein [28]

*Special relativity has led to the conclusion that inert mass is nothing more or less than energy, which finds its complete mathematical expression in a symmetrical tensor of second rank, the [stress- energy-momentum] tensor*

Therefore the mass of an object is fully and correctly described, not by a single number such as □□ but by the stress-energy-momentum tensor, $\mathbf{T}$. As an example, consider a rod lying at rest parallel to the x-axis in S whose unstressed proper mass density is $\square_0$ and whose proper volume is $V_0$. Such a system is referred to as an *open system*. Suppose also that the rod is under stress of magnitude $T_0^{xx}$ (the xx-component of $\mathbf{T}$ as measured in S) and that no work is done in S. The momentum density as measured in S′ is given by [29]

(6) $g = v\gamma^2\left(\rho_0 + T_0^{xx}/c^2\right)$

where $\mu_0 \equiv u_0/c^2$. Multiplying by the volume $V = V_0/\gamma$ gives the total momentum of the rod

(7) $p = gV = v\gamma^2\left(\rho_0 + T_0^{xx}/c^2\right)(V_0/\gamma)$
$= (\gamma\rho_0 V_0)v + \left(\gamma T_0^{xx} V_0/c^2\right)v$

Recall that the mass of the rod is the ratio of the rods momentum to its speed giving

(8) $m = p/v = \gamma\left(\mu + T_0^{xx} V_0/c^2\right)$

where $\mu \equiv \mu_0 V_0$. Notice that $m$ does not approach $\mu$ as $v \to 0$. If the rod were oriented perpendicular to its motion then this would not be the case. Therefore no unique value may be assigned to $\mu$ for a stressed rod. This is generally true for open systems. Similarly the inertial energy of the stressed rod can be shown to be

(9) $E = \gamma\mu c^2 + \gamma\beta^2 T_0^{xx} V_0$

In this case it is plainly seen that $E \neq \mu c^2$ contrary to claims made by the con-side. Therefore it is convenient to define a quantity in order to distinguish between $E/c^2$ and $p/v$. That quantity is $m$, (relativistic) mass. Since the value of $m$ depends on more than the magnitude of a proper mass and the rods speed then the assertion made by the pro-side of the debate that mass is not direction dependant is invalid in general. It should also be noted that, in general, the momentum of an object is does not necessarily point in the same direction the velocity points. While this is a well-established fact it is most likely a little known fact. [30] Only when an object is isolated (or is a closed system) will its momentum be parallel to its velocity.

III. MASS-DENSITY PARADOX

One is tempted to ignore the mass of an open system and only concern themselves with the mass of isolated objects/closed systems for which there is always a well-defined proper mass. How then should one define mass density using this viewpoint? In order to illustrate that certain problems may arise in such cases I will use as an example a homework problem posed in a recently published special relativity text. The student is given the strength of a magnetic field and asked to find the mass density. For the sake of argument assume that the magnetic field is static in S and there is no electric field in S either. Let the magnetic field point in the +y direction at the point **r**. We now determine the mass density of the magnetic field as measured in S'. Let $dV_0$ the volume of the "object," i.e. an infinitesimal volume element of the magnetic field, in S and $dV$ the volume of the object in S'. Let $dM_0$ be the mass as of the object as measured in S and $dM$ the mass of the object as measured in S'. The mass density in S' is then $\Box = dM/dV$. See in Fig. (1a)

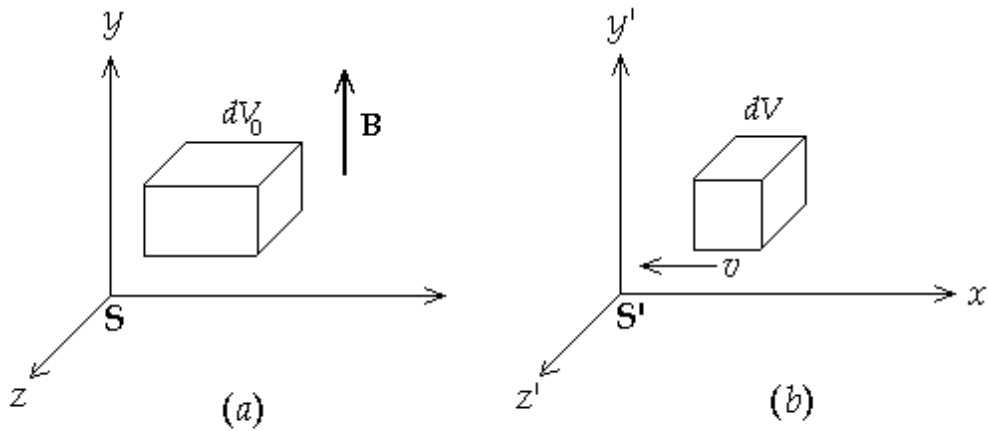

Fig. 1

The object is also shown in S' in Fig. (2b) where the object is moving in the $-x'$ direction. The magnitude of the field as measured in S is given by $\mathbf{B} = B\mathbf{e}_y$. Using the transformation rules for electric and magnetic fields we find that the value of the electric and magnetic fields in S' are found to be $\mathbf{E}' = \Box vB\mathbf{e}_z$ and $\mathbf{B}' = \Box B\mathbf{e}_y$ respectively. The momentum density $g$ is related to the Poynting vector **S**, defined as $\mathbf{S} \equiv \mathbf{E}\times\mathbf{B}/\Box$, by $g = \mathbf{S}/c^2$. In the present example the Poynting vector is found to be

(10) $\mathbf{S} = \dfrac{1}{\mu_0}(\gamma v B \mathbf{e}_z) \times (\gamma B \mathbf{e}_y) = -\gamma^2 v \left(\dfrac{B^2}{\mu_0}\right)\mathbf{e}_x = -\gamma^2 v u_0 \mathbf{e}_x$

The term in the parentheses is the energy density, $u_0$, as measured in S. Thus the magnitude of the momentum density $\mathbf{g}$ is given by $g = \rho_0 v$, where $\rho_0 \equiv u_0/c^2$. The magnitude of the infinitesimal momentum, $dP$, of the elemental object is then given by the product of the momentum density and the infinitesimal volume element $dV$ which yields

(11) $dP = \gamma v \rho_0 dV_0$

The mass element $dM$ is then given by $dP/v$ to obtain

(12) $dM = dP/v = \gamma \rho_0 dV_0$

The mass density as measured in S′ is then found to be

(13) $\rho = dM/dV = \dfrac{\gamma \rho_0 dV_0}{dV_0/\gamma} = \gamma^2 \rho_0$

The mass density in this case is a very familiar relationship in relativity. There is one factor of $\gamma$ for mass increase and another factor of $\gamma$ for volume decrease. However let us now orient the magnetic field parallel to the +x axis. The magnetic field as measured in S is now $\mathbf{B} = B\mathbf{e}_x$. The **E** and **B** fields now transform from S to S′ to give $\mathbf{B'} = \mathbf{B}$, $\mathbf{E'} = 0$. The Poynting vector is thus zero and therefore the momentum of the infinitesimal object is now zero. Thus the mass of the object in this frame is found to be zero for all speeds $v$. This result may come as a surprise to most since in this example there is rest energy without proper mass! The above results are quite interesting and *appear* to be a paradox. In S the object is considered to be at rest and has non-zero energy. In a frame moving perpendicular to the magnetic field the momentum of the object is non-zero and the mass density is as expected. If, however, we change the direction of the magnetic field we find that the moment is zero. Thus we cannot meaningfully assign a value of $m$, $\rho$ or $\mu$ unless we also specify the direction the object is moving. We certainly can't assign a value of $\mu$ since there is nothing intrinsic about the proper mass in this case. Is this truly a paradox, something not expected? The answer is *No!* In the first case we obtain the expected result due to the fact that the tension (i.e. stress < 0) in the magnetic field is perpendicular to the motion of the object and therefore the usual relationship for mass transformation will apply. However in the second case the field tension is parallel to the direction of motion and we therefore should not assume that we'd arrive at the same result that we did in the first example. In fact in the present case $T_0^{xx} = -u_0$ and when substituted into Eq. (7) yields zero momentum density. Mystery solved!

## III. PASSIVE GRAVITATIONAL MASS

Passive gravitational mass is sometimes referred to as "gravitational charge." In Newtonian mechanics this is the *m* in F = $GMm/r^2$ where M >> m is the active gravitational mass of a point source. In relativity all forces are velocity dependant. To determine what quantity fills this role in general relativity we will use the electromagnetic force as a guiding model. The components of the Lorentz 3-force are given in terms of the components of the Faraday $F^{\Box\Box}$ tensor as

(14) $f^k = qF^{k\nu}v_\nu$
    = (electric charge)×(components of EM field)×(velocity dependant terms)

where $v^\alpha \equiv (c, \mathbf{v})$, $v_\alpha \equiv g_{\alpha\beta}v^\alpha$. $g_{\Box\Box}$ are the components of the metric tensor. Eq. (14) holds for a point charge. The components of the gravitational 3-force are found to be [31]

(15) $f^k = m\Gamma^k_{\alpha\beta}v^\alpha v^\beta$
    = (gravitational charge)×(components of **G** field)×(velocity dependant terms)

Eq. (15) holds for a point particle. It is therefore relativistic mass *m* that plays the role of gravitational charge (passive gravitational mass) in general relativity. It is to be noted that $\gamma \equiv dt/d\tau$ no longer has the value $1/\sqrt{1-\beta^2}$. For a time orthogonal spacetime (i.e. $g_{0k} = 0$) it can be shown that, for a tardyon, *m* may be expressed in terms of the particle's proper mass as [32]

(16) $m(\Phi, v) = \dfrac{\mu}{\sqrt{1 + 2\Phi/c^2 - \beta^2}}$

where $\Phi \equiv c^2(g_{00} - 1)/2$. It follows from this analogy that the passive gravitational mass should be defined to be identical to the relativistic mass. Here we have a generalized form for the value of relativistic mass, valid even in non-inertial frames of reference (i.e. gravitational fields). Since $m(\Phi, 0) \neq \mu$ the terms *proper mass* and *rest mass* should not be used synonymously.

## IV. ACTIVE GRAVITATIONAL MASS

To examine the role relativistic mass plays as active gravitational mass in general relativity I start with Newtonian gravity as a guide. The active gravitational mass density is the $\rho$ in Poisson's equation. It is related to the gravitational potential $\Phi$ by

(17) $\quad \nabla^2 \Phi = 4\pi G \rho$

Einstein's equation

(18) $\quad \mathbf{G} = \dfrac{8\pi G}{c^2} \mathbf{T}$

is the equation in general relativity that determines the gravitational potentials $g_{\mu\nu}$. The potentials are embedded in the Einstein tensor **G**. To what extent does relativistic mass play in Eq. (18) when it is said that **T** describes it completely? I will answer this using the weak field limit for a relativistic fluid. For such a fluid the energy density is $u_0 = T^{tt}$ and $T_0^{xx} = T_0^{yy} = T_0^{zz} = p$ is the fluid pressure. Comparison with Eq. (8) shows that the mass density for such a fluid in frame S' is given by

(19) $\quad \rho = \gamma \left( \mu + p/c^2 \right)$

Taking the low speed limit and allowing for active gravitational mass contributions from the other two independent directions for velocity we find

(19) $\quad \rho = u_0/c^2 + 3p/c^2$

And when substituted into Eq. 17 becomes

(17) $\quad \nabla^2 \Phi = \dfrac{4\pi G}{c^2} (u_0 + 3p)$

which is identical to the weak field limit for a relativistic fluid when general relativity is applied. [33] This demonstrates how relativistic mass plays the role as active gravitational mass in general relativity. It is still to be noted that relativistic mass is completely described by **T**.

V. HISTORICAL BACKGROUND

In this section I walk the reader through an outline of the historical background that led up to the conclusions made above. In Einstein's 1905 paper on special relativity he assigned mass according to Mach's definition of mass. The following year Max Planck showed that the force on a charged particle could be expressed as

$$(21) \quad \mathbf{F} = \frac{d}{dt}(\gamma\mu\mathbf{v}) = \frac{d}{dt}(m\mathbf{v}) = q[\mathbf{E} + \mathbf{v}\times\mathbf{B}]$$

It was then natural to define momentum as the quantity $\mathbf{p} \equiv m\mathbf{v}$. Between the years 1906 and 1915 there appeared 6 notable papers regarding the concept of mass in relativity. [19, 20, 34-36] In 1906 Einstein assigned a mass density to the energy density of the electromagnetic field. In 1907 he demonstrated that the energy of a body was more than simply a function of the velocity of the body and $\Box c^2$ but also a function the stress acting on the body. During the years 1908 to 1912 Tolman and Lewis showed that one could assign velocity dependence to mass if one defines momentum as $\mathbf{p} \equiv m\mathbf{v}$ and if one demands momentum conservation in all inertial frames. By utilizing arbitrary collisions of two identical particles they arrived at $m = \Box\Box$. In Einstein's 1915 review article on general relativity he stated his conclusion that mass is described completely by the stress-energy-momentum tensor, **T**.

During the 20th century particle physics came to be a large part of physics and those studying particle physics used the theory of relativity to accomplish their work. The goal of particle physics is to study the intrinsic properties of particles such as proper mass and proper lifetime. Since particle physicists have no other use of the terms *mass* and *lifetime* the "proper" qualifier was dropped for the sake of convenience and simplicity. Thus the meaning of these terms took on a meaning different to that used by practitioners in general relativity. However since more physicists use special relativity than general relativity, pressure started to be applied within to the relativity community as a whole to define the term "mass" to mean proper mass. This pressure came, in a small extent, in the form of textbooks as well as journal articles.[3, 10] One comment that is often used to bolster this position is the following quote from Einstein in a letter to Lincoln Barnett dated June 19, 1948

*It is not good to introduce the mass* $M = m/\sqrt{1 - v^2/c^2}$ *of a body for which no clear definition can be given. It is better to introduce no other mass concept other than the "rest-mass" m. Instead of introducing M it is better to mention the expression for the momentum and energy of a body in motion.*

These words seem to say that Einstein was against the concept of relativistic mass. However this is a false conclusion. Had that been the case, Einstein would have stopped assigning a value to mass that depended on the gravitational potential as he did in his last version in *The Meaning of Relativity*, page 102 where Einstein writes [37]

*The inert mass is proportional to [ 1 + Φ ], and therefore increases when poderable masses approach the test body.*

This relation was derived by assuming small velocity. It was $m(□, 0)$ and *not* □ that Einstein was referring to here. This is the relation in the brackets in the above quote when an approximation is used and then divided by proper mass. The last edition was published in 1953, a full five years after his letter to Barnett. Einstein also attributed a value of mass to light in his and Infeld's text *The Evolution of Physics*. [18] I must admit that Einstein's inconsistency pertaining to the definition of mass is a mystery to me.

VII. CONCLUSIONS

I have demonstrated above that there are omissions, specious arguments and errors on both sides of the relativistic mass/proper mass debate that has left some areas in this dispute empty. I have also demonstrated that the claim that relativistic mass can be replaced by inertial energy is wrong in general. I gave one example and showed how such thinking could lead to an apparent paradox. I have also shown that the quantity which deserves the name *mass* more than anything else is relativistic mass since it is that quantity which possesses all three aspects of mass, i.e. inertial mass, passive gravitational mass and active gravitational mass. Accusations that relativistic mass is confusing cannot serve as a strike against relativistic mass since almost all aspects of relativity are confusing until they are mastered and once mastered they're not confusing at all. There is a certain consistency with calling the time component of 4-momentum *mass* and the magnitude of the same 4-vector *proper mass* since this is consistent with other 4-vectors and their magnitudes. I have labeled it here with a Greek symbol to reflect the fact that proper mass is an invariant and that invariant quantities in relativity are sometimes given Greek symbols. For example, the magnitude of a timelike 4-displacement is referred to as the *proper time interval* and labeled $d□$ whereas the timelike component is referred to as the *time interval* and labeled $d\tilde{□}$ Likewise, the magnitude of a spacelike 4-displacement is *proper distance* whereas the spacelike component is referred to as *distance*. Therefore one appends the term "proper" to the term given to a component these 4-vectors to obtain the name of the magnitude of that four vector. It would be desirable to have a unique meaning to all terms in physics but this seems unlikely to happen. For example, when one is working in quantum mechanics then the term *momentum* does *not* have the usual meaning as linear mechanical momentum but it is taken to mean *canonical momentum*. So too should we state the definition of the term "mass" and continue to use it consistently from therein within any relativity journal article or textbook. Practitioners in particle physics use the term mass to mean proper mass whereas those practitioners in general relativity might mean relativistic mass. So while the overall use of the term "mass" is proper mass it may be true that the majority of use in general relativity might use the term "mass" to mean relativistic mass or something else, like mass-energy (i.e. $m = E/c^2$).

APPENDIX - NOTATION

On notation; uppercase boldfaced letters, e.g. **P**, will denote either 4-vectors or 4-tensors while lowercase boldfaced letters, e.g. **p**, will denote 3-vectors. Lower case bold-faced letters with a tilde over them will represent 1-forms, e.g. the 1-form corresponding to 4-momentum will be represented by $\tilde{\mathbf{p}}$. Some important 4-vectors are the event 4-displacement $dX \equiv (cdt, dx, dy, dz)$, 4-velocity $\mathbf{U} = dX/d\tau$ where $d\tau$ is a proper time interval and 4-momentum $\mathbf{P} \equiv \mu \mathbf{U}$ where $\mu$ is the particle's *proper mass*. The components of a 4-vectors and 4-tensors are represented using a capital letter in italics with a Greek superscript, i.e. $P^\alpha$, $F^{\alpha\beta}$. I.e. $\mathbf{P} \equiv (P^0, P^1, P^2, P^3)$. The components of a 1-form will be represented using a capital letter in italics with a Greek subscript or subscript, i.e. $P_\alpha$ is be the $\alpha$ component of $\tilde{\mathbf{p}}$. By comparison with $dX$, $P^0$ is usually referred to as the *time-component* of **P** while the $P^i$ {as with all Latin indices i = 1, 2, 3} are referred to as the *spatial components* of **P**.

The term *scalar* as used here will be used to refer to a tensor of rank zero. Thus the scalar product of two 4-vectors will be a scalar. If one of the 4-vectors is considered a basis 4-vector then the scalar will be referred to as a *covariant* quantity otherwise it will be referred to as an *invariant* quantity. E.g. if **r** is the position 3-vector then $x = \mathbf{r} \cdot \mathbf{e}_x$, where $\mathbf{e}_x$ is a basis unit vector pointing in the $+x$ direction, is a covariant scalar whereas **r**•**r** is an invariant scalar. The term *covariant* used here is as defined in Lanczos. [38] Whenever frame dependant quantities such as "energy," "3-momentum" or "3-velocity" etc. are used there is an implied coordinate, i.e. the "observers frame of reference." To each inertial frame of reference in flat spacetime there is a time-like unit 4-vector $\mathbf{e}_0$ that corresponds to an observer at rest in said frame of reference. The 4-velocity corresponding to $\mathbf{e}_0$ will be labeled $\mathbf{U}_{obs}$. Thus the scalar product $\mathbf{P} \cdot \mathbf{U}_{obs}$ is to be considered a covariant quantity and not an invariant quantity.

The use of two inertial frames S and S′ in flat spacetime will be used below. S is comprised of a spatial Cartesian coordinates whose axes are labeled $x$, $y$, and $z$ whereas S′ is comprised of a spatial Cartesian coordinates whose axes are labeled

*x'*, *y'*, and *z'* and moving in the +*x* direction relative to S. S and S' so defined are said to be in *standard configuration*.